\documentclass[epj]{webofc}
\usepackage[utf8]{inputenc}
\usepackage[varg]{txfonts}   
\usepackage{booktabs}
\usepackage{xcolor}
\definecolor{darkred}{rgb}{0.4,0.0,0.0}
\definecolor{darkgreen}{rgb}{0.0,0.4,0.0}
\definecolor{darkblue}{rgb}{0.0,0.0,0.4}
\usepackage[bookmarks,linktocpage,colorlinks,
    linkcolor = darkred,
    urlcolor  = darkblue,
    citecolor = darkgreen]{hyperref}
\usepackage{subfigure}
\wocname{EPJ Web of Conferences}
\woctitle{Lattice2017}
\begin{document}
%
\selectlanguage{english}
\title{%
	Anderson localization in sigma models
}
\author{%
\firstname{Falk} \lastname{Bruckmann}\inst{1}\thanks{This work was supported by the DFG-grant BR 2872/6-1} \and
\firstname{Jacob} \lastname{Wellnhofer}\inst{1}\thanks{This work was supported by the DFG-grant BR 2872/7-1}\textsubscript{\tiny speaker} 
}
\institute{%
	Universität Regensburg, Universitätsstraße 31, 93053 Regensburg, Germany
}
\abstract{%
	In QCD above the chiral restoration temperature there exists an Anderson transition in the fermion spectrum from localized to delocalized modes. We investigate whether the same holds for nonlinear sigma models which share properties like dynamical mass generation and asymptotic freedom with QCD. In particular we study the spectra of fermions coupled to (quenched) CP(N-1) configurations at high temperatures. We compare results in two and three space-time dimensions: in two dimensions the Anderson transition is absent, since all fermion modes are localized, while in three dimensions it is present. Our measurements include a more recent observable characterizing level spacings: the distribution of ratios of consecutive level spacings.
}
\maketitle
\section{Introduction}\label{intro}

Anderson localization has been studied for a number of years in the context of QCD.
In the confined phase the QCD Dirac operator has no gap in the spectrum; 
its statistics is well described by the Gaussian unitary ensemble (GUE) of random matrix theory. 
Above the deconfinement temperature a gap emerges in the spectrum of the Dirac operator. 
At the same time the statistics of the low end of the spectrum becomes Poissonian, that is, 
the eigenmodes become localized \cite{GarciaGarcia:2006gr,Kovacs:2012zq,Nishigaki:2013uya,Giordano:2013taa,Ujfalusi:2015nha}, 
the same is true for the quenched case \cite{GarciaGarcia:2006gr,Kovacs:2017uiz}.
As nonlinear sigma models share a number of properties, 
such as asymptotic freedom and dynamical mass generation, with QCD, 
we will study the localization properties of CP(N-1) models in the following. 

In solid states physics Anderson localization describes the 
conductor-insulator transition. 
A regular lattice hamiltonian has Bloch eigenstates. 
These are extended modes and thus conducting. 
Adding a random on-site potential to emulate impurities in 
the conductor results in some modes being trapped/localized 
around these impurities.
A similar mechanism is at work in high temperature QCD. 
There the (local) Polyakov loop is $1$ nearly everywhere; 
however, there are defects where the Polyakov loop differs from $1$. 
Also here the low eigenmodes of the Dirac operator become trapped around these 
defects, see \cite{Bruckmann:2011cc}.
\section{Anderson localization}\label{anderson}

In order to study Anderson localization we can use a number of 
observables to distinguish localized from delocalized modes. 
One such observable is the participation ratio
\begin{align}
	PR(\lambda)
	&=
	\left(
	V
	\sum_x |\psi_\lambda(x)|^4
	\right)^{-1}, 
\end{align}
where $\psi_\lambda$ is a normalized eigenmode with eigenvalue $\lambda$. 
A localized mode $\psi$ is nonzero only in a small region around 
some $x_0$, ideally $\psi(x) = \delta_{x,x_0}$. 
In that case $PR \sim 1/V$, i.e., the participation ratio scales with the volume. 
In the other extreme of a completely delocalized mode, with constant $\psi(x) = 1/\sqrt{V}$
(or something similar), we have $PR \sim 1$, that is the participation ratio is 
independent of the volume. 

Another popular observable is the level spacing distribution $P(s)$. 
First the spectrum is unfolded, i.e., we rescale the eigenvalues 
$\lambda \to \bar\lambda$ such that the spectral density $\rho(\bar\lambda) = const.$. 
Then the level spacing $P(s)$ is defined as the probability of 
two adjacent eigenvalues being a distance $s$ apart, i.e., 
$\bar\lambda_{i+1} - \bar\lambda_{i} = s$.
For eigenvalues corresponding to localized modes 
the level spacing is Poissonian, 
\begin{align}
	\label{eq:ls_poisson}
P(s) = e^{-s}. 
\end{align}
For delocalized modes the level spacing obeys the usual Wigner-Dyson statistics;
in particular QCD and CP(N-1) are in the GUE: 
\begin{align}
	\label{eq:ls_GUE}
	P(s)
	&=
	\frac{32}{\pi^2} s^2 e^{-\frac{4x^2}{\pi} } 
\end{align}

In a similar spirit one can consider the distribution of the ratio of consecutive spacings, 
\begin{align}
r = \frac{\lambda_{i+1} - \lambda_i}{\lambda_i - \lambda_{i-1}} 
	\hspace{2em}\mathrm{or} \hspace{2em}
	\tilde r = \frac{\min(\lambda_{i+1} - \lambda_i,\ \lambda_i - \lambda_{i-1})}{\max(\lambda_{i+1} - \lambda_i,\ \lambda_i - \lambda_{i-1})} \in [0,1].
\end{align}
Also here localized modes can be distinguished by having Poissonian statistics, 
\begin{align}
	\label{eq:lsnu_poisson}
P(r) = 1/(1+r)^2
\end{align} 
as opposed to Wigner-Dyson statistics. 
In particular, for the GUE $P(r)$ is:
\begin{align}
	\label{eq:lsnu_GUE}
	P(r)
	&=
	\frac{4\pi}{81\sqrt{3}} 
	\frac{(r+r^2)^2}{(1+r+r^2)^4}.
\end{align}
The distributions for $r$ look similar to the ones for the level spacing, 
however, they have an algebraic decay for large $r$. 
The distribution of $\tilde r$ is very similar to the one for $r$, $P(\tilde r) = 2P(r)$
and it is confined to the interval $\tilde r \in[0,1]$. 
In our view the last point makes $P(\tilde r)$ more suitable for numerical studies, 
since increased statistics is directly added to the interval $[0,1]$, 
as opposed to the tail of the algebraic decay.
For more details on $P(r)$ and $P(\tilde r)$ see \cite{PhysRevLett.110.084101}.
\section{CP(N-1) models}\label{cpn}

The continuum action of CP(N-1) models 2 spacetime dimensions can be written
\begin{align}
	S
	&=
	\frac{1}{g} \int d^2x\, (D_\nu n)^\dagger (D_\nu n)
	&
	n^\dagger n &= 1,\  
	n\in \mathbb{C}^N
	\\
	&&
	\text{with}\quad 
	D_\nu 
	&=
	\partial_\nu + iA_\nu,
\end{align}
where $A$ is an auxiliary U(1) gauge field. 
In principle the gauge field can be (path-)integrated out 
because it has no kinetic term in the action, 
but we will keep it in the following. 

On the lattice we can replace the covariant derivative 
with a finite difference: $D_\nu n \to U_{x,\nu}n_{x+\hat \nu} - n_x$. 
With that we arrive at the lattice action, with $N\beta = 1/g$,
\begin{align}
	\label{eq:action_lat}
	S
	&=
	-N\beta \sum_{x,\nu} 
	\left(
	n^\dagger_x U_{x,\nu} n_{x+\hat\nu}
	+ \text{c.c.}
	- 2
	\right). 
\end{align}

For this study we produced gauge configurations with the action in eq. \eqref{eq:action_lat} 
for $N=4$ at various couplings $\beta$ and volumes.
In the following we use the U(1) gauge configurations
to study the spectra of the U(1) staggered Dirac operator, 
i.e., we study the quenched case.

\section{Results}\label{results}
\subsection{CP(N-1) in 2 dimensions}
\label{cpn2d}
\begin{figure}[h]
\begin{center}
    \includegraphics[scale=1,clip]{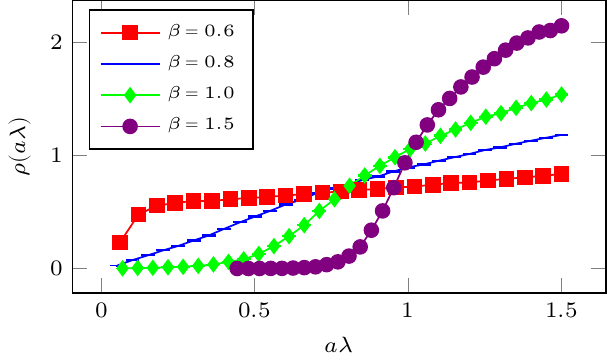}
\end{center}
	\caption{The spectral density of the Dirac spectrum 
	develops a gap at high temperature (large $\beta$).
	The simulation parameters were $N=4$, $N_t=4$, $N_s=128$. ($2d$ case.)}
\label{fig:eigdens}
\end{figure}

Also for CP(N-1) the spectrum of the Dirac operator develops a gap 
at high temperatures (large $\beta$), see figure \ref{fig:eigdens}. 
So it may well be that the statistics at the gap changes as well. 

\begin{figure}[p]
\begin{center}
		\includegraphics[scale=1.0]{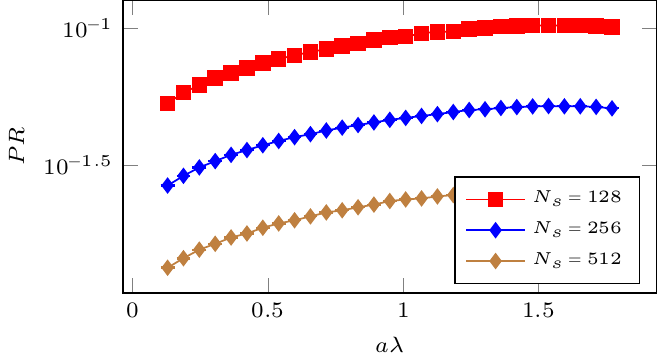}
		\includegraphics[scale=1.0]{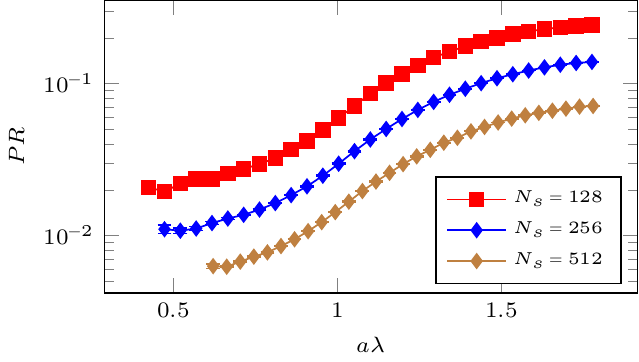}
\end{center}
\caption{($2d$ case.) The participation ratio as a function of the eigenvalue $a\lambda$.
	The left panel is for $\beta=0.6$, the right panel for $\beta=1.5$. 
	In both cases $PR\sim 1/V$ over the whole range of eigenvalues, 
	signaling localized modes only. }
\label{fig:pr2d}
\end{figure}
\begin{figure}[p]
\begin{center}
		\includegraphics[scale=1.0]{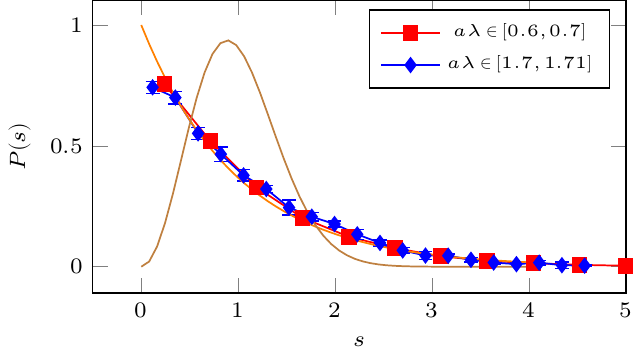}
		\includegraphics[scale=1.0]{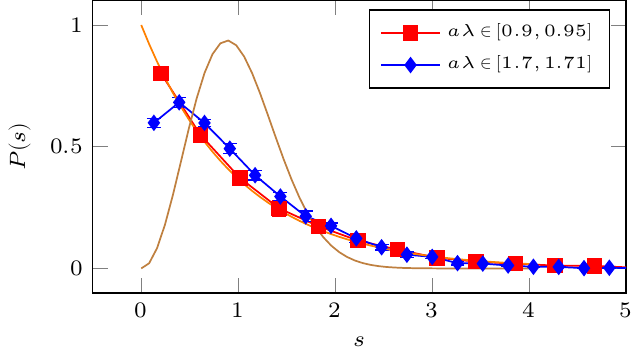}
\end{center}
\caption{($2d$ case.) The level spacing in different eigenvalue windows 
	for a lattice size of $4\times 512$. 
	The left panel is for $\beta=0.6$, the right one is for $\beta=1.5$.
	The distribution is Poissonian in all spectral windows. 
	For comparison we included the curves for pure Poissonian 
	and GUE statistics from eq. \eqref{eq:ls_poisson} and \eqref{eq:ls_GUE}.
	}
\label{fig:ls2d}
\end{figure}
\begin{figure}[p]
\begin{center}
		\includegraphics[scale=1.0]{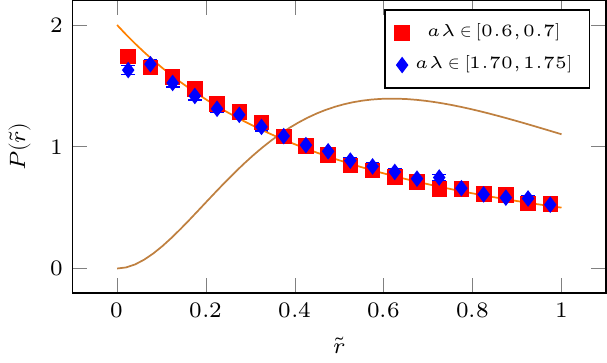}
		\includegraphics[scale=1.0]{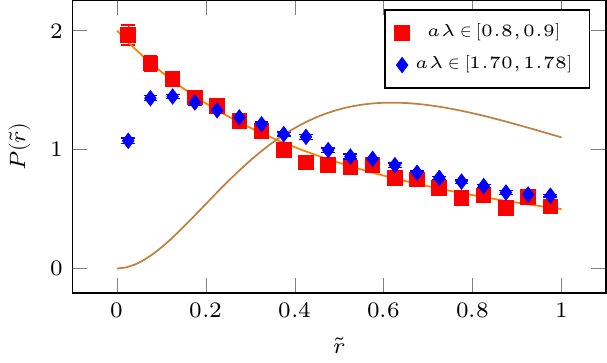}
\end{center}
	\caption{($2d$ case.) The ratio distribution $P(\tilde r)$ in different eigenvalue windows 
	for a lattice size of $4\times 512$. 
	The left panel is for $\beta=0.6$, the right one is for $\beta=1.5$.
	The distribution is Poissonian in all spectral windows. 
	For comparison we included the curves for pure Poissonian 
	and GUE statistics from eq. \eqref{eq:lsnu_poisson} and \eqref{eq:lsnu_GUE}.
	}
\label{fig:lsnu2d}
\end{figure}

First we look at the participation ratio in figure \ref{fig:pr2d}. 
We plot $PR$ as a function of the lattice eigenvalue $a\lambda$ 
for different volumes ($4\times 128$, $4\times 256$ and $4\times 512$) 
and temperatures ($\beta=0.6$ and $\beta=1.5$). 
We clearly observe scaling with the volume, $PR \sim 1/V$, 
across the whole range of eigenvalues. 
This indicates that all the eigenmodes are localized. 
Next we consider the level spacing, as well as the distribution of the ratio $\tilde r$, 
in figures \ref{fig:ls2d} and \ref{fig:lsnu2d}, respectively. 
The figures show Poissonian statistics for all the spectral windows, 
again indicating localized modes only. 

In the study of Anderson localization the dimensionality of the (single particle) model 
plays a crutial role. 
Namely, for $d<3$ the modes are always localized,
which is further confirmed by our study here. 

\subsection{CP(N-1) in 3 dimensions}
\label{cpn3d}
Since the two dimensional CP(N-1) model shows only localized modes, 
we want to study whether this changes in the three dimensional case. 

\begin{figure}[p]
\begin{center}
		\includegraphics[scale=1.]{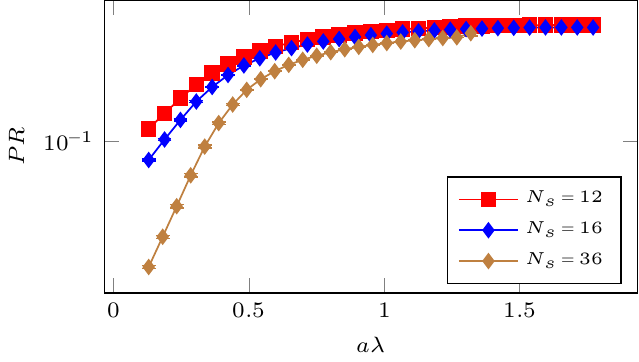}
		\includegraphics[scale=1.]{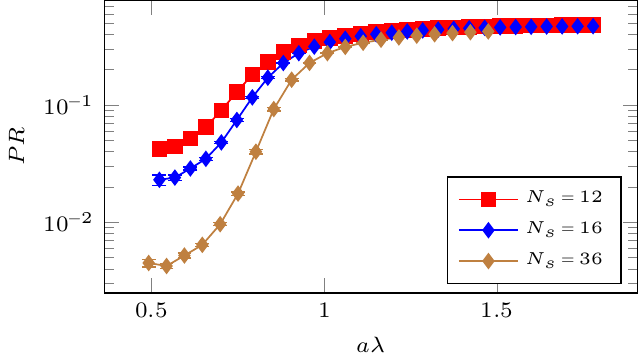}
\end{center}
\caption{($3d$ case.) The participation ratio as a function of the 
	eigenvalue $a\lambda$. 
	The left panel is for $\beta=0.6$, the right one is for $\beta=1.0$.
	The low eigenvalues clearly show volume scaling, 
	while $PR$ for the higher modes remains virtually unchanged 
	for diffenent volumes.
	This indicates that the low modes are localized while the 
	higher ones are spread out. }
\label{fig:pr3d}
\end{figure}

\begin{figure}[p]
\begin{center}
		\includegraphics[scale=1.]{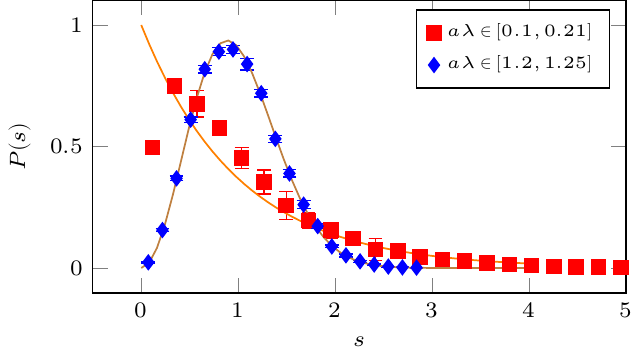}
		\includegraphics[scale=1.]{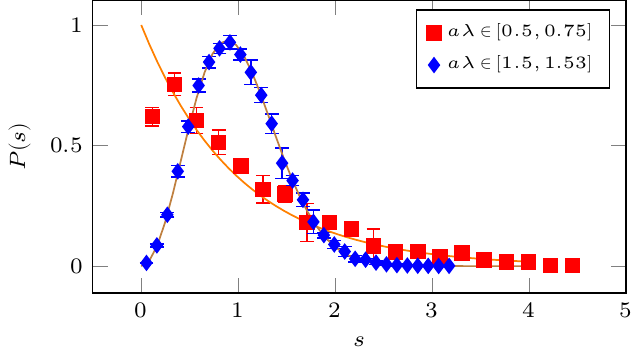}
\end{center}
\caption{($3d$ case.) The level spacing in different spectral windows 
	for a lattice size of $4\times 36\times 36$.
	The left panel is for $\beta=0.6$, the right one is for $\beta=1.0$.
	The low end of the spectrum shows Poissonian statistics, 
	while the higher modes are well described by the GUE. 
	For comparison we included the curves for pure Poissonian 
	and GUE statistics from eq. \eqref{eq:ls_poisson} and \eqref{eq:ls_GUE}.
	}
\label{fig:ls3d}
\end{figure}

\begin{figure}[p]
\begin{center}
		\includegraphics[scale=1.]{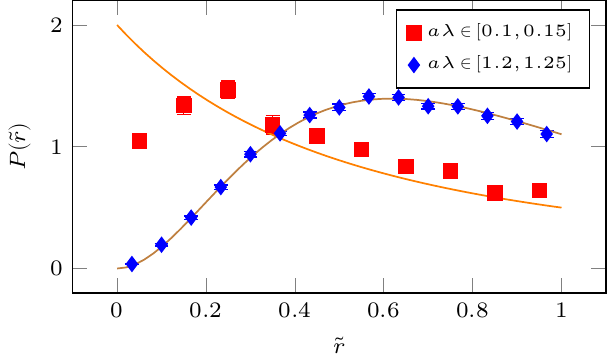}
		\includegraphics[scale=1.]{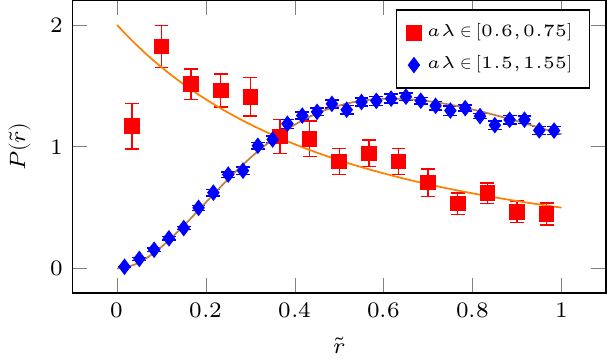}
\end{center}
\caption{($3d$ case.) The ratio distribution $P(\tilde r)$ in different spectral windows 
	for a lattice size of $4\times 36\times 36$.
	The left panel is for $\beta=0.6$, the right one is for $\beta=1.0$.
	The low end of the spectrum shows Poissonian statistics, 
	while the higher modes are well described by the GUE. 
	For comparison we included the curves for pure Poissonian 
	and GUE statistics from eq. \eqref{eq:lsnu_poisson} and \eqref{eq:lsnu_GUE}.
	}
\label{fig:lsnu3d}
\end{figure}

In figure \ref{fig:pr3d} we plot the participation ratio as a function of the 
lattice eigenvalue $a\lambda$ for different volumes and $\beta$s. 
We clearly observe that the $PR$ of the low modes scales with the volume, 
while for high modes the scaling is only very mild. 
This indicates that the low modes near the gap are localized. 
In figures \ref{fig:ls3d} and \ref{fig:lsnu3d} 
we again show the level spacing and the distribution of the ratio $\tilde r$, respectively. 
Also here the low modes show Poissonian statistics, 
while the higher modes are well discribed by the GUE, 
as indicated in the figures. 
This shows that, indeed, the $3d$ CP(N-1) model has an Anderson transition. 

\section{Conclusion}
\label{conclusion}
We have seen that CP(N-1) models also show an Anderson transition. 
However, this is only true in the three dimensional case (possibly also 
in higher dimensions). 
In two dimensions all modes appear to be localized, 
just as is the case in the Anderson model. 

In this sense an Anderson transition may be a more general feature 
of quantum field theories. 
In particular, 
the present work shows that having a gauged and gapped Dirac operator 
in $d>2$ is enough for an Anderson transition to appear. 
However, in order for this to have any physical relevance for the continuum theory 
the transition has to survive the continuum limit, as is the case in QCD. 
For CP(N-1) in three dimensions the situation is not so clear: 
There the coupling becomes dimensionful, 
indicating that the model is no longer renormalizable, 
at least perturbatively. 
Still there is the possibility that the CP(N-1) model in higher dimensions 
may be asymptotically safe. 
In \cite{Codello:2008qq} it is shown that for $d>2$ these models 
do have a nontrivial fixed point. 
So if CP(N-1) turns out to be asymptotically safe, 
then a continuum limit of our lattice approach could be feasable, 
even in higher dimensions. 

Another possibility is that there might be physical systems with a fixed (nonzero) lattice spacing, 
e.g., in solid state physics, 
which could be described by a CP(N-1) model. 
For such systems our analysis provides insights into the 
localization structure of fermions coupled to them. 
 
\paragraph{\bf Acknowledgements}
The authors thank Tilo Wettig for pointing out the 
observable in Ref. \cite{PhysRevLett.110.084101}.

\clearpage
\bibliography{lattice2017}

\end{document}